\documentclass[a4paper,fleqn,final]{cas-sc}

\usepackage[authoryear]{natbib}

\def\tsc#1{\csdef{#1}{\textsc{\lowercase{#1}}\xspace}}
\tsc{WGM}
\tsc{QE}

\begin{document}
\let\WriteBookmarks\relax
\def\floatpagepagefraction{1}
\def\textpagefraction{.001}

\shorttitle{}    

\shortauthors{}  

\title [mode = title]{Thermal-orbital evolution of Eris}  



\author[1]{Ryunosuke Akiba}[orcid=0000-0002-2681-3195]
\cormark[1]
\ead{rakiba@ucsc.edu}
\credit{Methodology, Software, Writing -- original draft,Writing -- review \& editing, Visualization, Investigation, Formal analysis}
\affiliation[1]{organization={Department of Earth and Planetary Sciences, University of California, Santa Cruz},
            addressline={1156 High Street}, 
            city={Santa Cruz},
            postcode={95064}, 
            state={CA},
            country={USA}}

\author[1]{Francis Nimmo}[orcid=0000-0003-3573-5915]
\ead{fnimmo@ucsc.edu}
\credit{Conceptualization, Methodology, Writing -- review \& editing, Funding acquisition}

\cortext[1]{Corresponding author}

\begin{abstract}
The large Kuiper Belt object (KBO) Eris is nearly as big as Pluto and has a small moon, Dysnomia. Constraints on the system’s spin and orbit characteristics were recently used to argue for a dissipative Eris, requiring a differentiated structure but not necessarily a subsurface ocean. Here, we model the thermal history of Eris coupled to its spin-orbital evolution, finding a subsurface ocean is preferred in order for Eris to be sufficiently dissipative. Spinning down Eris without an ocean is difficult, requiring a warm convecting ice shell protected by a thick insulating layer and very dissipative anelastic behavior in ice. Oceans make up 77-100\% of successful thermal-orbital evolution models, depending on the parameters assumed, which increases to $>$98\% when the Andrade $\beta$ parameter for ice is restricted to $\beta\leq3\times10^{-11}$~Pa$^{-1}$ s$^{-0.25}$. Oceans freeze over by the present day unless insulation (porosity, gas clathrates) or antifreeze are present. 
\end{abstract}

\begin{highlights}
\item Performed coupled thermal-orbital modeling of the Kuiper belt object Eris
\item A subsurface ocean is preferred to explain Eris's current orbital state
\item Oceans freeze completely by present day unless insulation or antifreeze are present
\end{highlights}

\begin{keywords}
Kuiper belt \sep 
Orbital evolution \sep 
Trans-neptunian objects \sep
Thermal histories \sep 

\end{keywords}
\maketitle

\section{Introduction and motivation}

Liquid water oceans outside Earth have been suggested to exist within various worlds throughout the solar system, from the icy satellites Europa and Enceladus to the dwarf planet Pluto \citep{nimmo2016ocean, castillo2022ocean}. The evolution of subsurface oceans is important for understanding the physical and chemical processes occurring in the interiors of planetary bodies, and the survival of these oceans is important for astrobiology due to their potential habitability \citep{hand2020habitability}. Icy dwarf planets are not expected to experience significant tidal heating, unlike the icy satellites orbiting giant planets. Candidate ocean worlds have been suggested among Kuiper belt objects (KBOs) \citep[e.g.,][]{hussmann2006subsurface, desch2009thermal, nimmo2016ocean}, with the oceans maintained by radioisotope decay, aided by the influence of volatiles and porosity on insulation and melting point \citep[e.g. Pluto,][]{kamata2019pluto}. Eris, a large KBO nearly as big as Pluto \citep{sicardy2011pluto} and the most massive dwarf planet known, is inferred from its synchronous spin to have had a dissipative past \citep{nimmo2023internal}, which can be explained by a warm, convecting ice shell or potentially a subsurface ocean. The existence and survival of a subsurface ocean on Eris would be additional evidence for another group of ocean worlds that can exist far from a host star without requiring the tidal energy provided by orbiting a giant planet. 

Characteristics of Eris and its moon Dysnomia are derived from telescope observations, which inform us about their bulk properties and orbital parameters (Table~\ref{tab:systemprops}). The surface of Eris is very bright with nearly uniform albedo; however, small variations have allowed for the determination of Eris's synchronous rotation \citep{szakats2023tidally, bernstein2023synchronous}. Eris may be maintaining its high albedo against darkening by radiation by atmospheric collapse due to temperature variations experienced during its eccentric orbit \citep{sicardy2011pluto} or by renewal of the surface by seasonal volatile transport \citep{hofgartner2019ongoing}. Volatile retention modeling suggest CH$_4$, N$_2$, and CO ices are supported at the surface \citep{schaller2007volatile}, and spectroscopic observations have detected CH$_4$ and N$_2$ \citep{grundy2024measurement}. The D/H ratio  of surface methane ices as measured by JWST is found to be lower than that of primordial methane, suggesting internal production by geochemical processes \citep{grundy2024measurement, glein2024moderate}.

Eris has the highest bulk density of the observed KBOs of $\rho = 2.43 \pm 0.05$~g~cm$^{-3}$ \citep{holler2021eris}, implying a composition with high rock fraction $>$0.8 \citep{bierson2019using, sicardy2011pluto} and a thin ice shell. For a differentiated rock-ice body, $\rho_{rock}=3500$ kg m$^{-3}$ gives a rock-ice ratio of 0.85/0.15 and an ice shell thickness of $\sim$180~km. Eris's high bulk density may be due to a giant impact origin of the Eris-Dysnomia system between differentiated bodies that removed a significant fraction of ice \citep[e.g.,][]{arakawa2019early, brown2023masses}. A more organic-rich, comet-derived composition \citep{mckinnon1997composition,castillo2023compositions, arakawa2025low} would add low density organics to the rocky core and decrease the fraction of ice. Recent analysis of \textit{Gaia} photometry by \cite{ortiz2025analysis} identified a peak in periodicity ($\sim$18.85 h) different from the previously detected peak at Dysnomia's orbital period (378.862 h) \citep{szakats2023tidally, bernstein2023synchronous}. \cite{ortiz2025analysis} suggest this period may correspond to the rotation period of Eris or of an additional, undiscovered close-in satellite. This would affect Eris's present-day orbital state and lower its bulk density after subtracting the satellite mass. We neglect this possibility in this study, and assume the synchronous rotation period independently determined by \cite{szakats2023tidally} and \cite{bernstein2023synchronous}. 

Key observations of the Eris-Dysnomia system being doubly synchronous \citep{szakats2023tidally, bernstein2023synchronous} and an upper bound on the mass of Dysnomia \citep{brown2023masses} allowed \cite{nimmo2023internal} to place a lower bound on how dissipative Eris is, providing a strong constraint on Eris's interior structure. For Dysnomia to have spun down Eris over 4.5 Gyr or less, Eris was inferred to have been quite dissipative, requiring a differentiated structure but not necessarily a subsurface ocean. However, \cite{nimmo2023internal} did not model the thermal evolution of Eris. The thermal evolution of the body determines how a convecting ice shell and/or ocean can evolve and influence, through tides, its orbital evolution. Coupled thermal-orbital modeling is a powerful tool to gain insight into the internal structure and thermal history of planetary bodies from observed orbital properties \citep[e.g.,][]{meyer2010coupled, hammond2024triton}, especially for some of the most distant objects in the solar system \citep[e.g.,][]{bagheri2022tidal,arakawa2025low}. To understand what internal structures are consistent with Eris's spun-down state, we investigate the long-term thermal history of Eris coupled to its orbital evolution. We conclude that subsurface oceans are possible on Eris and are likely required to explain Eris's current orbital state, though they may not have survived to the present. 

The rest of this paper is organized as follows. We describe the thermal model, which tracks the initiation of convection in the ice shell and/or formation of an ocean. We relate the thermal state to tidal dissipation, which couples the thermal and orbital evolution calculations. We then use the thermal-orbital model to find conditions that successfully reproduce Eris's current orbital state. 

\begin{table}[width=.9\linewidth,cols=4,pos=h]
\caption{System properties. We use the central values in our models.}\label{tab:systemprops}
\begin{tabular*}{\tblwidth}{@{} LLLL@{} }
\toprule
Parameter & Variable & Value & Reference\\
\midrule
System mass (kg) & $M$ & $(1.6466\pm 0.0085)\times 10^{22}$ & \cite{holler2021eris}\\
Mass ratio upper bound () & $f$ & 0.0084 & \cite{brown2023masses}\\
Semi-major axis (km) & $a$ & $37273\pm 64$ & \cite{holler2021eris}\\
Spin/orbit period (hour) & $P$ & 378.862 & \cite{holler2021eris}\\
Eris radius (km) & $R_p$ & $1163\pm 6$ & \cite{sicardy2011pluto}\\
Dysnomia radius (km) & $R_s$ & $350\pm 58$ & \cite{brown2018medium}\\
\bottomrule
\end{tabular*}
\end{table}

\section{Methods}

\subsection{Thermal evolution}

We adopt a differentiated interior model that consists of a hydrosphere and rocky core, where the hydrosphere can evolve into an ice shell and subsurface ocean depending on the thermal state. We model its thermal evolution numerically, solving a 1-D heat equation using a finite difference method, considering conductive and convective modes of heat transport while allowing for an ocean to form and evolve. Long-lived radioisotopes in the rocky core provide a heat source, and solid-state convection in the ice shell is modeled using a mixing length theory (MLT) parameterization \citep{kamata2018one}. The heat transport equation --- including conductive and convective heat flux terms --- in the ice shell and core is as follows:
\begin{equation}
\begin{split}
     \rho C_p \frac{\partial T}{\partial t} &= \frac{1}{r^2}\frac{\partial}{\partial r} \left( r^2 F_{cond} + r^2 F_{conv} \right) + Q(t), \\
     F_{cond} &= -k \frac{\partial T}{\partial r}, \\
     F_{conv} &= -k_v \left[ \frac{\partial T}{\partial r} - \left(\frac{\partial T}{\partial r}\right)_{ad} \right],
\end{split}
    \label{eq:heat_equation}
\end{equation}
where $\rho$ is density, $C_p$ is specific heat, $k$ is thermal conductivity, $Q(t)$ is the volumetric radiogenic heat production term, and $\left(\frac{\partial T}{\partial r}\right)_{ad}$ is the adiabatic temperature gradient. Values for the material properties of the hydrosphere and core layers are given in Table~\ref{tab:materialprops}. The rocky core is assumed to be conductive ($F_{conv} = 0$) and heated by the radioactive decay of long-lived radioisotopes, nominally assuming elemental abundances of a CI chondrite composition \citep{turcotte2002geodynamics} (Supplementary Table 1). The decay of short-lived $^{26}$Al is not considered. The abundances of radioisotopes in Eris are unclear, and more cometary compositions have been suggested in KBOs \citep{mckinnon1997composition,castillo2023compositions, arakawa2025low}, which would introduce more organics and decrease the concentrations of radioisotopes and lower the heating rate. We later consider a case with reduced heating (Section~\ref{sec:organic}). 

\begin{table}[width=.9\linewidth,cols=4,pos=h]
\caption{Material properties used for thermal-orbital evolution}\label{tab:materialprops}
\begin{tabular*}{\tblwidth}{@{} LLLL@{} }
\toprule
Parameter & Variable & Value & Unit\\
\midrule
Density of ice & $\rho_{ice}$ & 930 & kg m$^{-3}$\\
Thermal expansivity of ice & $\alpha$ & 1.56$\times10^{-4}$ & K$^{-1}$\\
Rigidity of ice & $G_{ice}$ & 3.3 & GPa\\
Latent heat of fusion & $L_{ice}$ & 333 & kJ kg$^{-1}$\\
Activation energy & $E_a$ & 60 & kJ mol$^{-1}$\\
Reference viscosity & $\eta_{ref}$ & $10^{13} - 10^{15}$ & Pa s \\
Andrade $\beta$ parameter & $\beta$ & $10^{-12} - 10^{-10}$ & Pa$^{-1}$s$^{-0.25}$ \\
Density of ocean & $\rho_{ocean}$ & 1000 & kg m$^{-3}$\\
Density of rock & $\rho_{rock}$ & 2800-3500 & kg m$^{-3}$\\
Thermal conductivity of rock & $k_{rock}$ & 4 & W m$^{-1}$ K$^{-1}$\\
Specific heat of rock & $C_{p,rock}$ & 1000 & J kg$^{-1}$ K$^{-1}$\\
Viscosity of rock & $\eta_{rock}$ & $10^{22}$ & Pa s \\ 
Rigidity of rock & $G_{rock}$ & 30 & GPa \\
Surface temperature & $T_s$ & 35 & K \\
\bottomrule
\end{tabular*}
\end{table}

In the ice shell, temperature-dependent properties of ice are considered. Thermal conductivity and specific heat of water ice Ih are given by (\cite{andersson2005thermal} and \cite{choukroun2010thermodynamic}, respectively)
\begin{equation}
    \begin{split}
        k_{ice}(T) &= \frac{632}{T} + 0.38 - 0.00197 T, \\
        C_{p,ice}(T) &= 74.11 + 7.56 T.
    \end{split}
    \label{eq:iceprops}
\end{equation}
Thermal conductivity values for a pure-water ice shell can vary from $\sim 2$~W~m$^{-1}$~K$^{-1}$ near the melting point to $\sim 18$~W~m$^{-1}$~K$^{-1}$ at the surface temperature $T_s$=35~K \citep{sicardy2011pluto}, resulting in a mean value of $\sim 5$~W~m$^{-1}$~K$^{-1}$. 

The viscosity of ice strongly affects convective heat transfer and the viscoelastic response to tides. An Arrhenius relation is used to describe the temperature-dependent ice viscosity assuming Newtonian behavior 
\begin{equation}
    \eta(T) = \eta_{ref} \exp{\left[\frac{E_a}{R_g}\left(\frac{1}{T}-\frac{1}{T_{ref}}\right)\right]},
    \label{eq:viscosity}
\end{equation}
where $E_a$ is activation energy, $R_g$ is the gas constant, and a reference viscosity $\eta_{ref}$ at $T_{ref}=270$~K is estimated to be in the range $10^{13}$ and $10^{15}$~Pa~s \citep{goldsby2001superplastic}. The strong temperature dependence causes cold surface viscosity values to be very large, often forming a thick lithosphere.

The convective heat flux term $F_{conv}$ is estimated with an ``effective'' thermal conductivity $k_v$ controlled by a modified characteristic mixing length parameter defined by Eq.~5 of \cite{kamata2018one}. This approach of parameterizing convection in icy bodies with and without oceans has been applied to long-term evolution models of icy satellites and Pluto \citep{kamata2019pluto, kimura2020stability}, the methods of which we closely follow. The ice shell is assumed to be conductive ($F_{conv} = 0$) until solid-state convection is initiated, following a Rayleigh number calculation. The Rayleigh number for an ice shell with temperature-dependent viscosity is defined 
\begin{equation}
    Ra = \frac{\rho g \alpha \Delta T d^3}{\kappa \eta_b},
\end{equation}
with gravity $g$, thermal expansivity $\alpha$, temperature contrast across the layer $\Delta T$, ice shell thickness $d$, thermal diffusivity $\kappa$, and viscosity at the base $\eta_b$. Solid-state convection in the ice shell can initiate when $Ra$ exceeds the critical Rayleigh number
\begin{equation}
    Ra_c = 20.9 (x p)^4,
\end{equation}
where $x$ depends on the spherical geometry of the system and $p=\gamma \Delta T$ quantifies the viscosity difference across the layer. For a spherical shell with a core radius fraction of the range we explore for Eris, around 0.85-0.9, a reasonable fit to 3-D numerical convection models gives $x=1.35$ \citep{robuchon2011thermal}. For a temperature-dependent ice viscosity (Eq.~\ref{eq:viscosity}), $\gamma=E_a R_g^{-1} T_b^{-2}$ for temperature at the base of the ice shell $T_b$. As the initially conductive ice shell is warmed by heat transferred from the core, $\eta_b$ decreases and $Ra$ increases, to the point where convection may initiate. 

An ocean can form if temperatures at the rock-ice interface exceed the melting point, calculated from a pure water ice phase diagram using the SeaFreeze package \citep{journaux2020holistic}, which gives $T_m=265$~K for pure water ice at pressures beneath a 120~km ice shell on Eris. The ocean temperature is assumed to be uniform at the melting point at the ice-ocean interface; temperatures at the core surface and base of the ice shell are pinned at $T_m$ for Eq.~\ref{eq:heat_equation} upon formation of an ocean. The change in ocean thickness then depends on the difference in incoming heat flux from the core and outgoing heat flux through the base of the ice shell and the latent heat of fusion for water. The change in ice shell thickness $D$ is described by the energy balance \citep{kamata2018one},
\begin{equation}
    \rho_{ice} L_{\mathit{eff}} \frac{d D}{d t} = F_{out} - F_{in}, 
    \label{eq: ocean thickness}
\end{equation}
for outgoing and incoming heat fluxes, $F_{out}$ and $F_{in}$, respectively, at the ice-ocean interface. $F_{in}$ is the conductive heat flux from the top of the core transferred across the ocean. The effective latent heat $L_{\mathit{eff}}$ \citep[Eq.~30 in ][]{kamata2018one} accounts for the temperature of the ocean as the pressure of the ice-ocean interface, and thus $T_m$, changes. 

Eris is initiated as a differentiated, two-layer, cold, isothermal ice-rock body, heated from within the core ($Q(t)$) and Eq.~\ref{eq:heat_equation} is solved with a finite difference method to evolve through time. Boundary conditions are given by a fixed surface temperature $T_s = 35$~K and zero thermal gradient at the center. The initial isothermal temperature chosen (100~K) does not greatly affect the long-term evolution unless set close to the surface temperature, which can sometimes prevent the body from warming due to the large thermal conductivity of ice at low temperatures predicted by Eq.~\ref{eq:iceprops}. Higher initial temperatures can allow earlier convection and ocean formation, and we check its effect on results in Section~\ref{sec:pureice}. 

The radial temperature profile is discretized with node spacing $\Delta r = 5000$~m for the core, $\Delta r = 1500$~m at the top 10\% of the core, and $\Delta r_{ice} = 800$~m for the ice shell. Radial step sizes were chosen with consideration of stability, computational efficiency, and testing for smooth behavior of the timing of convection and/or ocean formation across parameter space. Ocean formation partially decouples the heat equation (Eq.~\ref{eq:heat_equation}) with two constant temperature boundary conditions $T=T_m$ at the top and bottom of the ocean. The ocean is allowed to grow continuously, with ice nodes adjusting by slightly expanding or contracting before being discarded, aiming to keep $\Delta r_{ice}$ relatively constant. A special case is made for a convecting ice shell over an ocean, which for stability can compress node spacing up to half ($\Delta r_{ice} > 400$~m). 

The discretized temperature profile is evolved in time with an explicit method, with time step $\Delta t$ adaptively controlled by the Courant-Friedrichs-Lewy condition. We set the conductive time step, $\Delta t_{cond} = 0.3 \Delta r_{ice}^2 \max(\kappa_{ice})^{-1}$ for the largest value of thermal diffusivity $\kappa$ encountered in the ice shell. If convection is relevant, $\Delta t_{conv} = 0.25 \Delta r_{ice}^2 \max(\kappa_{v})^{-1}$, and the smaller time step of the two is used.

\subsection{Insulation of the ice shell by porosity}

Ice porosity and its evolution through pore closure may be an important component of the history of KBOs in explaining their bulk densities \citep[e.g.,][]{mckinnon2008structure, prialnik2008growth, bierson2019using} and in their insulating effects on thermal evolution \citep[e.g.,][]{besserer2013convection, bierson2018implications}. Porosity of the ice shell may be primordial from accretion, or to a lesser degree generated from impacts \citep[e.g.,][]{arakawa2002impact}. We explore the coupled thermal-porous evolution of the ice shell following the methodology of \citep{besserer2013convection}. The thermal conductivity is governed by 
\begin{equation}
    k_{ice,p}(T, \phi) = k_{ice}(T) \left( 1 - \frac{\phi}{\phi_p}\right)^{a\phi + b},
\end{equation}
where $k_{ice}(T)$ is given by Eq.~\ref{eq:iceprops}, $\phi$ is the porosity, $\phi_p = 0.7$ is the percolation limit, and $(a,b) = (4.1,0.22)$ are conductivity factors \citep{shoshany2002monte}. As the thermal structure in the ice shell evolves, viscous flow removes porosity following \citep{fowler1985mathematical, nimmo2003origins}
\begin{equation}
    \frac{\partial \ln{\phi}}{\partial t} = -\frac{P(z)}{\eta(T)},
    \label{eq:poro}
\end{equation}
where $P$ is overburden pressure at that depth. As a porous ice shell warms, the pores close and become less insulating, only surviving in the near-surface. A warm initial state of the ice shell can close pores as it cools, so the initial porous layer thickness implicitly depends on the initial temperature. For the assumed initial temperature of 100 K, initial porosity can span the entire ice shell, but will disappear for an initial temperature of 200 K. Porosity is discretized at the same radial nodes as temperature and evolved by Eq.~\ref{eq:poro}. For simplicity we neglect the evolution of the bulk density due to the lower density porous ice, consistent with some other coupled thermal-porosity evolution modeling of Pluto and other KBOs \citep{bierson2018implications, bierson2019using}, and assume a constant radius of Eris. Eris's high density implies a high rock-ice ratio and a relatively thin ice shell whose total thickness is less affected by porosity. 

Other insulating mechanisms (gas clathrates) have been suggested to sustain an ocean on Pluto \citep{kamata2019pluto} (Section \ref{sec:clathrate}). In these cases, a simpler approach is taken of assigning constant low thermal conductivity values at the nodes corresponding to the location of the insulating layer.  

\subsection{Spin-orbit evolution}

Driven by tidal torques, the orbit of Dysnomia evolved away from Eris as Eris spun down to eventually reach a synchronous state \citep{szakats2023tidally, bernstein2023synchronous}. To model this spin-orbital evolution, we use the formulation of \cite{cheng2014complete}, assuming that eccentricities are zero throughout and neglecting higher-order spin-orbit resonances. A nonzero present-day eccentricity for Dysnomia was reported by \cite{holler2021eris}. \cite{nimmo2023internal} discussed this possibility and concluded that dissipation in Eris could conceivably have increased Dysnomia's eccentricity; we do not consider this issue further. We assume that Dysnomia spun down quickly due to its small size relative to Eris and is synchronous throughout the orbital evolution. With these assumptions, the rotation rate of Eris $\Omega_i$ and semi-major axis $a$ evolve by 
\begin{equation}
\begin{split}
    \frac{d \Omega_i}{d t} &= - \frac{3 G M_j^2}{2 C_i} \frac{k_{2,i}}{Q_i} \frac{R_i^5}{a^6} [\rm{sgn}(\Omega_i - n)], \\
    a^{-1} \frac{d a}{d t} &= 3n \frac{k_{2,i}}{Q_i} \frac{M_j}{M_i} \left(\frac{R_i}{a}\right)^5 [\rm{sgn}(\Omega_i - n)],
\end{split}
\label{eq:orbit}
\end{equation}
where $M$ is mass, $R$ is radius, $n$ is mean motion, $C$ is the polar moment of inertia, and $Q/k_2$ describes the tidal response of the body. The subscripts $i,j$ refer to the primary (Eris) and secondary (Dysnomia), respectively (Table~\ref{tab:systemprops}). For initial conditions after Dysnomia formation, we assume an initial separation of 7~Eris radii, corresponding to a rotation period of Eris of 31 hours, assuming angular momentum conservation. Results of the thermal-orbital evolution are not sensitive to this parameter, except for very large separations (Supplementary Section 2). 

Coupling between the numerical integration of Eqs.~\ref{eq:orbit} and the thermal evolution arises through the changing tidal response, $Q/k_2$, of Eris as its internal viscoelastic structure changes. Here, $k_2$ is the tidal Love number and $Q$ is the dissipation factor, where a low $Q/k_2$ indicates a dissipative body. For a given internal structure of Eris, the tidal Love number is computed with the open-source California Planetary Geophysics Code written in Python \citep{ermakov_2024_14029354, akiba2022probing, park2025io}. The code uses the method of \cite{takeuchi1972seismic} to calculate tides and closely follows the implementations of \cite{kamata2015tidal, martens2016using, martens2019loaddef} (see Appendix A of \cite{akiba2022probing} for more details and Supplementary Section 4 for relevant benchmarks). Eris is assumed to be incompressible and dissipation in the ocean is not modeled. The tidal response depends on the viscoelastic structure of Eris (Table~\ref{tab:materialprops}) and forcing frequency $\Omega_i - n$. A Maxwell rheology and constant viscosity is assumed for the rocky core while the more dissipative ice shell is described by a more realistic Andrade $\beta$ rheology model \citep[e.g.,][]{mccarthy2012planetary, nimmo2023internal, bierson2024impact} and temperature-dependent viscosity (Eq.\ref{eq:viscosity}). An Andrade beta parameter range of $\beta=10^{-10} - 10^{-12}$~Pa$^{-1}$~s$^{-n}$ is taken, with $n=0.25$ \citep{mccarthy2016tidal}. Expressions for the rheology models are listed in Appendix~\ref{sec:appdxrheo}. 

Tidal Love numbers are computed from the thermal evolution using the internal structure and ice shell temperature profiles output at timesteps of 20 Myr, with denser output during ocean formation. An ice shell viscosity profile is computed, then downsampled to 20 constant viscosity layers to define a viscoelastic structure for the tidal Love number calculation. At very high ice viscosities for cold ice shells, the rheology is dominated by the anelastic (Andrade) term and $Q/k_2$ mostly depends on the chosen value of $\beta$. The rocky core remains too cold to be dissipative, thus the tidal response is controlled by the temperature-dependent viscosity structure of the ice shell and is more dissipative for warm, low viscosity ice shells and in the presence of an ocean, which decouples the ice shell from the rocky core. Porous ice is treated the same as pure ice for computing the tidal response. Thin-ocean resonances ($<100$ m) can cause the body to be extremely dissipative \citep{kamata2015tidal}, but are short-lived and require a $\leq 100$~m scale global ocean that is not grounded. Thus, we only compute tides with oceans $>500$~m thick, taking a conservative approach to the dissipative effect of thin oceans. 

Although orbital energy is dissipated as tidal heating in Eris, its contribution to its thermal evolution is small due to the low mass of Dysnomia as the tide-raising body. Comparing the orbital energy of the assumed initial configuration to the present-day state, total tidal dissipation amounts to $\sim10^{25}$~J. Over 4.5 Gyr, the average tidal dissipation rate of $\sim0.07$~GW is small compared to the radiogenic heating rate in the silicate core, which averages $\sim 150$~GW. Unlike the icy satellites of the Jovian or Saturnian systems, there is no large tide-raising body to cause dissipation and maintain an ocean after formation. The lifespan of an ocean, if formed, is mostly controlled by the heat input from the silicate core and the efficiency of heat transport through the ice shell \citep[e.g.,][]{robuchon2011thermal}. Thus, we neglect tidal heating from orbital evolution, allowing for a partial decoupling of the thermal-orbital evolution, where thermal models can be run independently before applying the orbital calculation. 

\section{Results}

To match current observations of the Eris-Dysnomia system, thermal-orbital evolution must spin down Eris within 4.5 Gyr. To do so requires thermal evolution to produce internal viscoelastic structures that are sufficiently dissipative. For each model run, Eris is assigned a rock density (and thus ice-rock ratio and ice layer thickness), an ice reference viscosity, and an Andrade $\beta$ parameter value from a range (Table~\ref{tab:materialprops}) and assigned a mode of ice shell insulation. A range of rock densities of $\rho_{rock} = 2800-3500$ kg m$^{-3}$ is used to cover potentially organic-rich compositions, corresponding to ice fractions of 6.5 - 15\% or hydrosphere thicknesses 70-180~km. The thermal evolution is computed, then the output viscoelastic structures are used to calculate tides for the spin-orbit evolution, which is run until the present-day state is reached (a successful simulation) or 4.5~Gyr has elapsed (unsuccessful simulation).

\subsection{Individual thermal-orbital evolution runs}

\begin{figure*}
	\centering
	  \includegraphics[width=\textwidth]{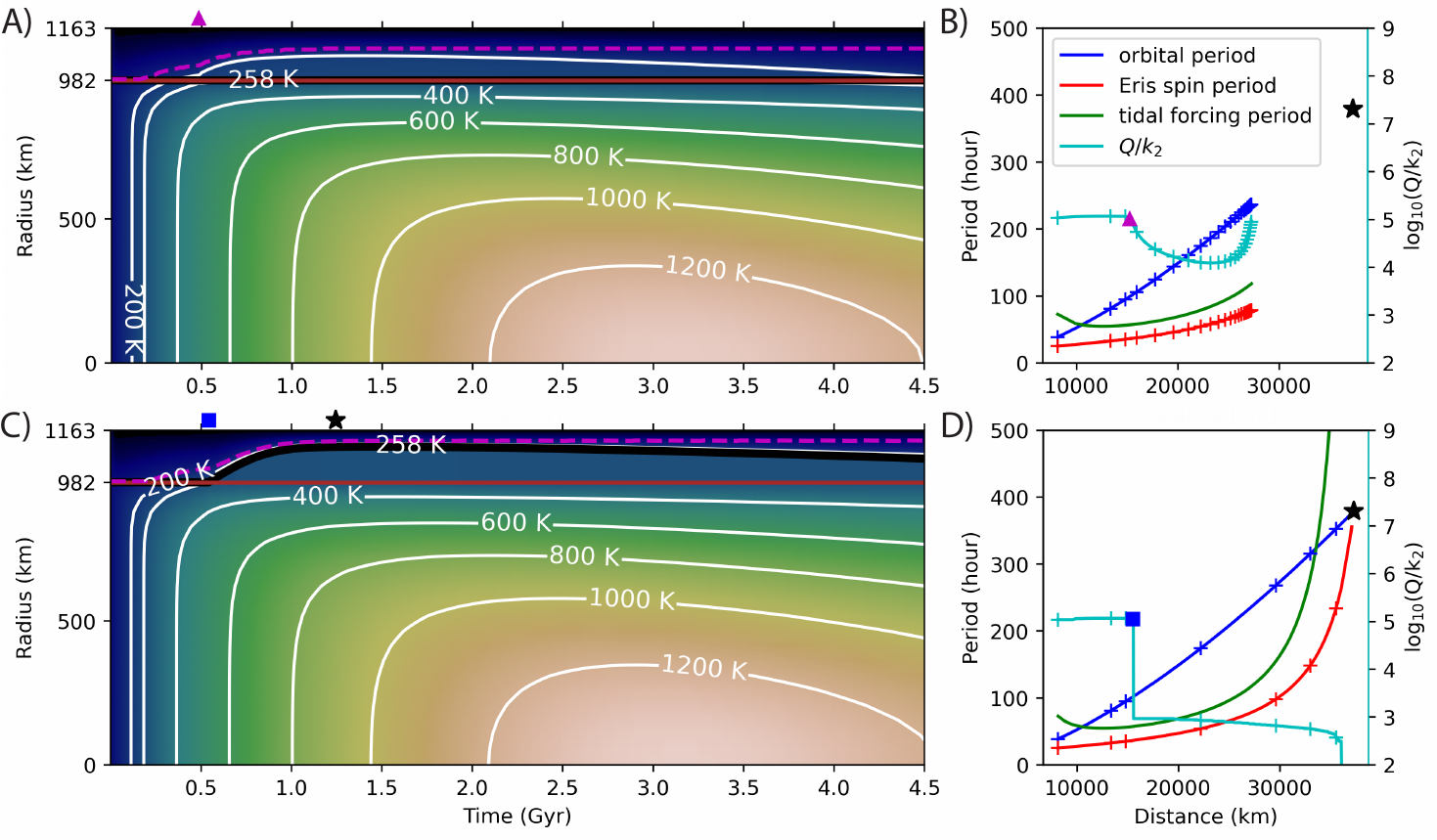}
	\caption{Two examples of thermal-orbital evolutions of Eris for low (A,B) and high (C,D) reference viscosity, leading to convection and ocean formation, respectively. In this run, $\rho_{rock}=3500$~kg~m$^{-3}$, Andrade $\beta=10^{-12}$~Pa$^{-1}$~s$^{-0.25}$, and $\eta_{ref}=10^{13}$ (A,B) or $\eta_{ref}=10^{15}$ Pa~s (C,D). Thermal evolutions (A,C) are colored and contoured with temperatures. The horizontal red line is the top of the silicate core, the dashed magenta line indicates the extent of the porous layer, and the black line (C) is the top of the ocean. Orbital evolution (B,D) starts at initial separation of 7 Eris radii, crosses are at intervals of 200 Myr, and the present-day state is indicated by the star. The cyan axis plots the tidal response on a log scale. The timing of initiation of convection, ocean formation, and reaching of the current state are indicated in all plots as the triangle, square, and star, respectively. Panels A,B and C,D depict unsuccessful and successful simulations, respectively.}
	\label{fig:EvolutionExample}
\end{figure*}

Before exploring the parameter space, we highlight two individual cases. Figures~\ref{fig:EvolutionExample}A,C show the thermal evolution of a two-layer Eris with a thick hydrosphere (180~km, normalized moment of inertia of 0.328) for low (A) and high (C) reference viscosities ($\eta_{ref}=10^{13},10^{15}$~Pa~s, respectively). In both cases, Eris is heated from within the silicate core by radiogenic heating and insulated by an initially porous ice shell that evolves as temperatures rise and pores are viscously closed, thinning the porous layer (dashed magenta curve). The core temperatures reach a maximum temperature of 1300~K at the center while the bulk of the core cools after 2 Gyr. In the ice shell, the low viscosity case (A) exceeds the critical Rayleigh number and initiates convection just before 500 Myr \citep[cf.][]{robuchon2011thermal}. The formation of a warm convective layer thins the porous layer until the point of strongest convection around 1.2~Gyr, corresponding to peak radiogenic heat flux from the core. At later times, the porous layer thickness stays constant.  

Efficient convective heat extraction in the low viscosity case (A) prevents the ice shell from heating up to the melting point ($T_m\sim 258$~K) unlike the high viscosity case (C), where a 100+km thick ocean (black curve) forms beneath a conductive ice shell starting around 600 Myr \citep[cf.][]{robuchon2011thermal}. Porosity evolves along the conductive temperature gradient, thinning with the ice shell as the ocean thickens. The ocean starts to refreeze following peak radiogenic heat flux, however in this case the porous ice layer is sufficiently insulating to maintain the subsurface ocean to the present day. 

Following the two thermal histories of Eris presented, the corresponding orbital evolutions are shown (Figure~\ref{fig:EvolutionExample}B,D) assuming an Andrade ice rheology with $\beta = 10^{-12}$~Pa$^{-1}$~s$^{-0.25}$ and evolving towards the present-day state (star). In the low viscosity case (B), the tidal response $Q/k_2$ (cyan curve) decreases upon the start of convection (triangle) as the ice shell becomes more dissipative. However, this convecting ice shell without an ocean is not sufficiently dissipative to spin down Eris. Although a dissipative convecting layer is formed, the overlying cold, conductive lid is thick, contributing to a high value for $Q/k_2$. Similarly, a purely conductive ice shell without an ocean cannot despin Eris. 

The formation of an ocean decouples the ice shell from the rigid interior for the high viscosity case (D), making Eris much more dissipative and spinning it down rapidly. The tidal response, $Q/k_2$ (cyan curve), drops by two orders of magnitude at ocean formation and the dissipative Eris reaches the present-day state by 1.3 Gyr, representing a successful simulation. A frequency effect is seen as $Q/k_2$ decreases in response to an increasing forcing period (green curve) as Eris approaches its synchronous state. The decoupling of the ice shell and rocky interior makes Eris sufficiently dissipative. As seen in the following sections, the formation of an ocean largely guarantees a successful simulation, except for a purely Maxwell rheology case, where decoupling can be insufficient. 

\subsection{Spinning down Eris without insulation}
\label{sec:pureice}
\begin{figure*}
	\centering
	  \includegraphics[width=\textwidth]{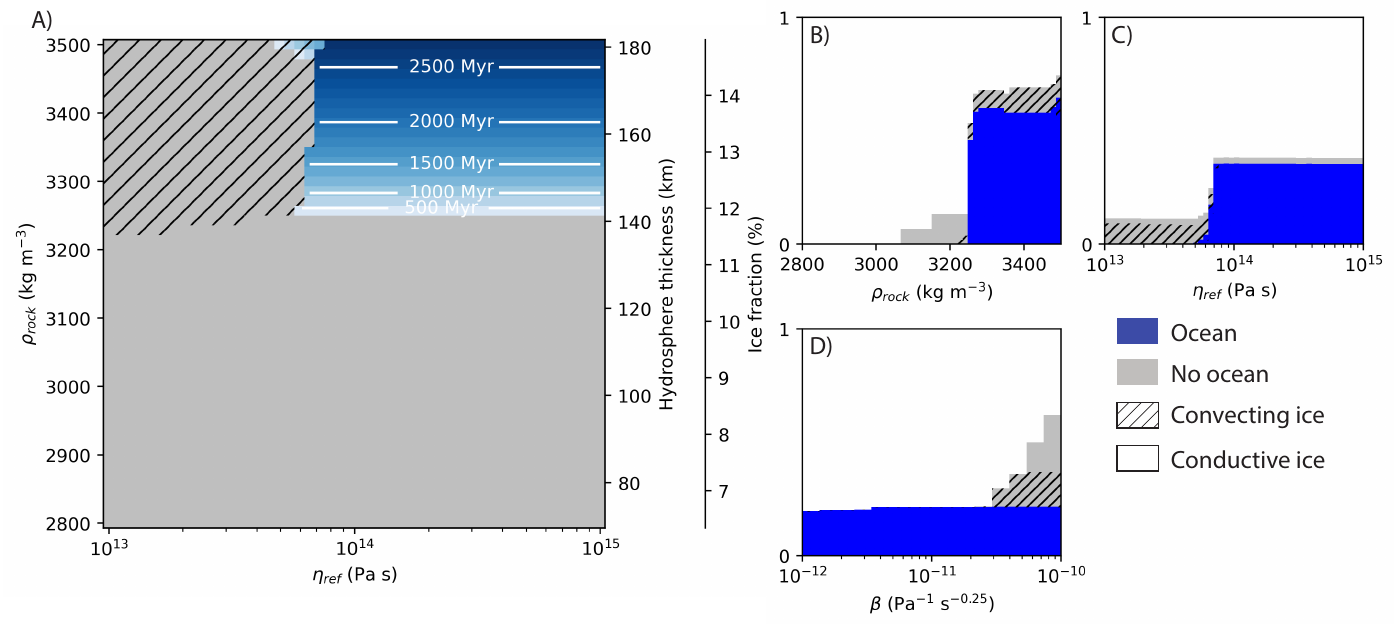}
	\caption{Simulation results assuming a pure ice shell of thermal evolution (A) and thermal-orbital evolution (B,C,D) over a grid of ice reference viscosity $\eta_{ref}$, core density $\rho_{rock}$, and Andrade $\beta$ parameter. All oceans freeze over by the present day, and the duration over which they were sustained is contoured (A). (A) is not constrained by orbital evolution, while (B,C,D) plot histograms of successful thermal-orbital evolution simulations. Models that produce an ocean are shaded blue and where a convecting ice shell exists (with or without an ocean) is hatched. 27\% of all tested cases shown in (A) are successful and 77\% of these have oceans.}
	\label{fig:pureice}
\end{figure*}

The result of the parameter space exploration for a pure water ice, non-porous case are summarized in Figure~\ref{fig:pureice}, with the full space of thermal evolutions (A) for combinations of rock density $\rho_{rock}$ and ice reference viscosity $\eta_{ref}$, and the subset of successful simulations (B,C,D) shown as histograms with an additional variable of the Andrade beta parameter $\beta$. Thermal evolution modeling (A) finds oceans forming for high $\rho_{rock}$ (thick hydrospheres) when ice viscosity is high, where conductive ice shells cannot remove enough heat to balance the heat flux from the core (see Figure~\ref{fig:EvolutionExample}C). None of these oceans survive to the present day, but they are sustained for a time that ranges from a few hundred Myr for thinner hydrospheres to 2.5 Gyr for the thickest. Convection can initiate in thick, low viscosity ice shells, with a small subset in which convecting ice shells overly an ocean. For the set of convecting ice shells overlying an ocean (hatched blue areas), the efficient heat extraction by convection freezes the ocean within 1 Gyr of formation. However, the majority of the explored space, with rock density $<3200$ kg m$^{-3}$ ($<140$ km hydrosphere thickness), consists of conductive ice shells that remain relatively cold throughout the history of Eris due to a lack of insulation of the ice shell (mean $k_{ice}\gtrsim 5$~W~m$^{-1}$~K$^{-1}$). 

When the constraint of a spun-down Eris is applied, the subset of successful simulations represent 27\% of all cases tested, of which around 77\% have oceans, 13\% a convective ice shell without an ocean, and 10\% a conductive ice shell without an ocean (Figure~\ref{fig:pureice}B,C,D). All cases with a long-lived ocean ($>500$~Myr) are found to be successful; the decoupling of the ice shell and rocky interior makes Eris sufficiently dissipative to successfully spin down for all values of $\beta$. Some short-lived ($<500$~Myr), thin ($<1$~km) oceans do not spin down Eris for low $\beta$ values, behaving similarly to the no-ocean cases. Short-lived ocean cases with higher $\beta$ values do spin down Eris, however, this is mainly due to a highly dissipative ice shell rather than the result of decoupling. This may suggest that a transient, thin ocean of metamorphic origin ($<10$ km) from dehydrating rock between 550 and 850~K \citep{courville2023timing} will not be able to spin down Eris unless there is a way to maintain the ocean (e.g., by antifreeze or gas clathrates). A warmer initial temperature of the ice shell by 50~K slightly expands the convecting ice and ocean forming regions to thinner hydrospheres, but the effect on the thermal-orbital evolution is small, resulting in $\sim3$\% more successful simulations (Supplementary Section 3). 

There were no successful cases with $\rho_{rock}<3050$~kg~m$^{-3}$, corresponding to a hydrosphere thinner than 110 km or an ice fraction less than 9\% (Figure~\ref{fig:pureice}B). Because most of the dissipation comes from the ice rather than the rocky core, it is difficult to successfully spin down with a thin hydrosphere. This may limit the amount of low density, organic-rich material that Eris could contain. However, more critically, spinning down Eris without an ocean requires a high $\beta$ value ($>2.5\times10^{-11}$~Pa$^{-1}$~s$^{-0.25}$) (Figure~\ref{fig:pureice}D). Here, the anelastic component of the tidal response provides a higher ``baseline'' $Q/k_2$ value when the ice shell is cold and the viscoelastic component is small. Thus, in these cases the orbital evolution is less sensitive to the thermal state of the ice shell, and the highest $\beta$ value allows even a ``cold'', conductive Eris to spin down with a sufficiently thick ice shell. When the ice shell is warmer due to convection, slightly lower $\beta$ values can successfully spin down Eris to the present state. We discuss the role of $\beta$ further below (Section~\ref{sec:icerheo}).

\subsection{Porous ice insulation}
\label{sec:resultporo}

\begin{figure*}
	\centering
	  \includegraphics[width=\textwidth]{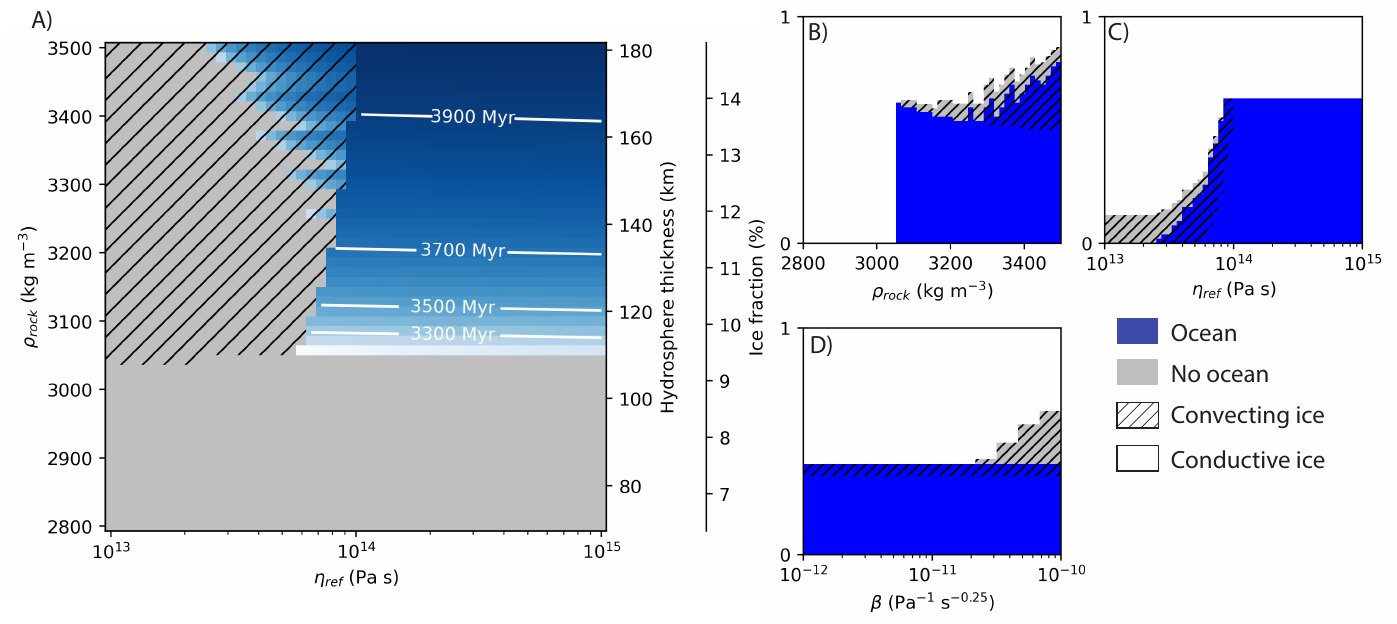}
	\caption{Simulation results starting with a porous ice shell of thermal evolution (A) and thermal-orbital evolution (B,C,D) over a grid of ice reference viscosity $\eta_{ref}$, core density $\rho_{rock}$, and Andrade $\beta$ parameter. All oceans survive to the present day, and their ages are contoured in the conductive region (A). Ages are not contoured in the convective region. (A) is not constrained by orbital evolution, while (B,C,D) plot histograms of successful thermal-orbital evolution simulations. Models that produce an ocean are shaded blue and where a convecting ice shell exists (with or without an ocean) is hatched. 44\% of all tested cases are successful and of these 90\% have oceans. The sawtooth pattern between convective ice shell with ocean and without ocean (A) is an artifact of the discretization of porosity affecting the convective heat flux.}
	\label{fig:poroice}
\end{figure*}

The results of parameter space exploration for the case for ice shells that evolve from an initially porous state are shown in Figure~\ref{fig:poroice}, representing an upper bound on the insulation that can be provided by porosity. The thermal evolution (A) shows a wider range of rock densities (thinner hydrospheres) that support convective ice shells and oceans than the previous pure ice case (Figure~\ref{fig:pureice}A), resulting from the insulation provided by the porous ice layer. Although the porous layer thins as the ice shell warms (Figure~\ref{fig:EvolutionExample}A,C), it provides sufficient insulation to maintain oceans to the present day, with lifetimes ranging from 3 to 4 Gyr. Porosity evolution (Eq.~\ref{eq:poro}) causes the ice shell to be slightly more insulating for higher viscosities, as pores close more slowly. This results in non-horizontal contours of ocean age (Figure~\ref{fig:pureice}A), but is a small effect (<100 Myr difference for a given $\rho_{rock}$). 

The region of convective ice shells over oceans also expands --- where convection initiates but viscosity is high enough that convective heat transport still allows the base of the ice to melt. The higher heat flux through the ice shell causes these oceans to form later than the conductive cases for the same hydrosphere thickness. Present-day ocean thicknesses range from 20 to 80~km largely depending on $\rho_{rock}$ for conductive ice shells and are thinner with a convective ice shell (Supplementary Figure 4). The boundary between convective ice shells with and without oceans exhibits a sawtooth pattern due to the discretization of the porous layer, which affects the thermal conductivity in the ice shell, and thus the convective heat flux. 

When the constraint of a spun-down Eris is applied, the subset of successful simulations (Figure~\ref{fig:poroice}B,C,D) represents 44\% of total cases tested, consisting of around 90\% with oceans and 10\% with a convective ice shell without an ocean (Table~\ref{tab:runsummary}). A larger fraction of the explored space is successful compared to the pure ice case, largely due to the higher frequency of ocean formation. However, even with insulation from porosity, the low $\rho_{rock}<3050$~kg~m$^{-3}$ cases are excluded because the hydrosphere gets too thin to be both warm and dissipative. No conductive ice shells without an ocean are able to despin Eris by the present day. The conductive ice shells without an ocean for this case are thinner than the uninsulated case and are thus less dissipative in general. If an ocean is present, conductive models can reproduce the current orbital state. A convective ice shell alone can spin down Eris, but similar to the uninsulated case, these successful convective ice shells require more dissipative ice ($\beta >2.5\times10^{-11}$~Pa$^{-1}$~s$^{-0.25}$) and cannot spin Eris down for low $\beta$ values. We discuss the effects of different ice rheology models in Section~\ref{sec:icerheo}. 

\begin{table*}[cols=4,pos=h]
\caption{Summary of simulation runs. The fraction of successful simulations across the parameter space of $\rho_{rock}, \eta_{ref}, \beta$ is listed in ``\% successful.'' These successful cases are split into ocean and no ocean (with conductive or convective ice shells), and of the ocean cases, the fraction that survive to the present day are in parentheses. For a more constrained case of the Andrade $\beta$ parameter $\beta\leq3\times10^{-11}$~Pa$^{-1}$s$^{-0.25}$, almost all simulations require an ocean.}\label{tab:runsummary}
\begin{tabular*}{\tblwidth}{@{} LLLLLLL@{} }
\toprule
Run description & \% successful & \multicolumn{4}{c}{Distribution of successful simulations} & If $\beta\leq3\times10^{-11}$ \\
\cline{3-6}
 &  & \multicolumn{2}{c}{No ocean} & \multicolumn{2}{c}{Ocean (Survives)} & fraction of successes\\
\cline{3-4} 
 &  & Convective & Conductive &  &  & with ocean \\
\midrule
Pure ice & 27.2\% & 13.2\% & 9.4\% & 77.4\% & (0.0\%) & 99.9\% \\
Porous ice & 44.2\% & 10.0\% & 0.0\% & 90.0\% & (100\%)& 99.9\%  \\
Antifreeze & 70\% & 0.0\% & 0.0\% & 100\% &  (100\%)& 100\% \\
Basal clathrates (10 km) & 69.6\% & 0.9\% & 0.0\% & 99.1\% & (8.7\%)& 100\% \\
Surface clathrates (10 km) & 78.4\% & 3.5\% & 0.0\% & 96.5\%  & (28.5\%)& 98.5\% \\
Surface clathrates (5 km) & 54.2\% & 7.6\% & 0.0\% & 92.4\%  & (15.5\%)& 99.9\% \\
Porous ice, Maxwell rheology & 20.9\% & 0.0\% & 0.0\% & 100\%  & (100\%)&  \\
Porous ice, Andrade $\zeta$ rheology & 39.9\% & 0.0\% & 0.0\% & 100\% & (100\%) &  \\
Porous ice, 20\% less heating & 25.1\% & 17.5\% & 14.8\% & 67.7\% & (100\%) & 100\%  \\
Porous ice, 30\% less heating & 14.8\% & 13.6\% & 56.4\% & 30.0\% & (100\%) & 100\% \\
\bottomrule
\end{tabular*}
\end{table*}

\section{Discussion}

\subsection{Effects of antifreeze}

The presence of antifreeze compounds such as ammonia can significantly lower the melting temperature and support oceans at lower temperatures \citep{choukroun2010thermodynamic}. Ammonia has been observed in spectra of KBOs \citep[e.g.,][]{brown2000evidence, barucci2008surface} and their presence on the surface may hint at past cryovolcanism \citep{desch2009thermal, neveu2015prerequisites}. An ocean composition of 10~wt\% ammonia depresses the melting point to $T_m\sim249$~K from $T_m=263$~K at 0.1~GPa \citep{choukroun2010thermodynamic, kimura2020stability}. We test the effects of antifreeze (e.g., ammonia) by shifting the ice-water phase curve by -20~K for a pure water ice shell without insulation. In this case, thermal-orbital models are successful across 70\% of the parameter space, all of which are composed of oceans which are sustained to the present day (Table~\ref{tab:runsummary}) and with $\rho_{rock}>3020$~kg~m$^{-3}$. There is no sensitivity to viscosity, as the low temperatures in the ice shell prevent convection \citep[cf.][]{kimura2020stability}; an ocean forms before temperatures can rise enough to initiate convection. Although not modeled here, the concentration of an antifreeze changes as ocean volume evolves, exhibiting its maximum effect as the ocean freezes. This may result in a picture similar to Figure~\ref{fig:pureice} but with more oceans surviving to the present day. Salts (e.g., NaCl, MgSO$_4$, etc) would also contribute to a depression of the melting point, but not as dramatically as ammonia.

\subsection{Clathrates and location of insulation}
\label{sec:clathrate}

Clathrate hydrates (clathrates) are a solid cage of water molecules trapping gas molecules, which form as a volatile-rich ocean crystallizes \citep[e.g.][]{courville2023timing, castillo2019conditions, kamata2019pluto}. We focus on methane clathrates, with methane produced by water-rock reactions, which are insulating with thermal conductivity up to an order of magnitude lower than that of pure water ice \citep{castillo2011ceres,castillo2023compositions}. The potential location of clathrate layers varies based on Eris's early history and differentiation, with the thickness and timing of a freezing primordial ocean dictating whether and where a clathrate layer can form \citep{courville2023timing}. Later thermal metamorphism can release volatiles and help preserve an ocean by adding insulation in the form of clathrates. We test two main cases of clathrate insulation following \cite{kamata2019pluto}: a 10~km basal layer of thermal conductivity 0.6 W m$^{-1}$ K$^{-1}$ and a surface layer with thermal conductivity 1.0 W m$^{-1}$ K$^{-1}$ with thicknesses 5 and 10 km. No porosity is assumed for the ice shell. We note that a full treatment of clathrate insulation requires thermal-geochemical modeling starting before Eris's differentiation, similar to \cite{courville2023timing}, but this is out of the scope of our work. We assume that the clathrate layers exist by the time Dysnomia forms and orbits at a distance of 7 Eris radii, and that thicknesses do not change subsequently. 

Clathrate insulation returns the highest fraction of successful simulations for the explored parameter space of $\rho_{rock}$, $\eta_{ref}$, and $\beta$ at 70\% for basal clathrates, and 78\% for surface clathrates of 10 km thickness (Table~\ref{tab:runsummary}, Supplementary Figure 5). Basal clathrates enhance ocean formation and preserve them enough to spin down Eris, but most oceans freeze over by the present day (leaving 9\% of simulations with present-day oceans). A consequence of basal clathrates is that a large part of the ice shell remains cold and suppresses convection, which makes up just 1\% of successful models. This cold ice shell results in a high effective thermal conductivity despite the basal insulation, and cannot sustain most oceans to the present day. 

In contrast, a surface layer of clathrates insulates and warms the entire ice shell, enhancing both ocean formation and convection \citep{kamata2019pluto} which helps successfully spin down Eris. This effect is seen in the 10~km surface clathrate case, which allows oceans to form and spin down Eris for the lowest values of $\rho_{rock}$ tested (2800 kg m$^{-3}$)--- the only case to do so. This is because the clathrate lid provides large effective thermal insulation early in Eris's thermal history, more than what porosity can provide initially (Eq.~\ref{eq:poro}). However, as the ice shell warms, thermal conductivity for porous ice scales with temperature, which helps preserve more oceans to the present day than for the clathrate case (Table~\ref{tab:runsummary}). Additionally, this is the only case that has simulations that can spin down Eris via a convecting ice shell without an ocean and the smallest value of Andrade $\beta = 10^{-12}$ Pa$^{-1}$ s$^{-0.25}$. A thinner 5~km surface clathrate layer also enhances ocean formation and convection, but to lesser effect. The 5~km surface clathrate case requires $\beta > 1.3\times10^{-11}$ Pa$^{-1}$ s$^{-0.25}$. 

\subsection{Differences between ice rheology models}
\label{sec:icerheo}

Different ice rheology models have been used to describe stress-strain relationships and the viscoelastic behavior of ice in response to periodic forcing. The Maxwell model is the most simple but lacks anelastic responses. Andrade models include an anelastic term and come in different flavors, $\beta$ and $\zeta$ \citep[e.g.,][]{bierson2024impact} to better match experimental data. While the $\beta$ model approaches an anelastic response at high viscosity, the $\zeta$ model includes a dependence on viscosity (Supplementary Figure 6). As ice viscosity increases, the $\zeta$ model eventually becomes less dissipative than the $\beta$ model. The present study primarily uses the Andrade $\beta$ model. We tested the effects of Maxwell and Andrade $\zeta = 10^{-2} - 10^{1}$ models on the orbital evolution of the porous ice case (same thermal models as Figure~\ref{fig:poroice}). Definitions of the rheology models are listed in Appendix~\ref{sec:appdxrheo}. The Maxwell rheology is less dissipative, with 21\% of the parameter space successfully spinning down Eris (Table~\ref{tab:runsummary}). No simulations without an ocean succeed, and of those with oceans, low viscosity ice shells are required to be sufficiently dissipative, excluding most cases with $\eta_{ref}>2\times10^{14}$ Pa s (Supplementary Figure 7). The Andrade $\zeta$ model results in an Eris sufficiently dissipative with an ocean and decoupled ice shell, but is unable to spin down Eris without an ocean.

The range of $\beta$ chosen for the present study of $\beta = 10^{-12} - 10^{-10}$~Pa$^{-1}$ s$^{-0.25}$ is conservative, with the upper bound of $10^{-10}$~Pa$^{-1}$ s$^{-0.25}$ lying outside experimentally determined values of $10^{-12} - 3\times10^{-11}$~Pa$^{-1}$ s$^{-n}$ \citep[see Figure 4 in][]{bierson2024impact}. This has a large impact on the statistics presented in Table~\ref{tab:runsummary} (see last column), as many of the successful no-ocean simulations require the highest values of $\beta$ to be sufficiently dissipative. With an upper bound of $\beta =3\times10^{-11}$~Pa$^{-1}$ s$^{-0.25}$, almost no non-ocean simulation can spin down Eris to the present-day orbital state, except some cases insulated with a 10~km thick surface lid of clathrates.

Dissipation in the silicate core is small compared to the ice shell with the assumed Maxwell rheology. An Andrade rheology for the silicates would be more dissipative \citep[e.g.,][]{renaud2018increased, bierson2024impact} and could allow more successful no-ocean cases. However, in our view silicate dissipation is less likely to be important in the case of Eris because its rocky core is cold ($<$1300~K) compared to the mantles of Earth or Io, and has a thick lithosphere ($\sim 300$ km with temperatures $<800$ K for Fig.~\ref{fig:EvolutionExample}) that resists tidal deformation. While the uppermost part of the lithosphere may be more dissipative due to porosity or alteration, this dissipative layer thickness will be much smaller than any such layer at Enceladus  \citep{roberts2008tidal, choblet2017powering}, owing to the higher pressures ($\sim100$~MPa) below the hydrosphere. Similarly, since cooling-induced cracking likely only extends to $6$ km \citep{vance2007hydrothermal}, dissipation in this thin layer is unlikely to be comparable to that of the ice shell. 

\subsection{Organic-rich compositions}
\label{sec:organic}

The present study assumed a CI chondritic composition of the rocky core to model radiogenic heating. However, KBOs have been suggested to be composed of more comet-like, organic-rich compositions \citep{castillo2023compositions, arakawa2025low, truong2024broad}, which would decrease the concentration of radioisotopes. We test cases of 20\% and 30\% less long-lived radiogenic heating (Table~\ref{tab:runsummary}), where the 30\% less case is slightly more than the effect of depleting potassium by a factor of 2 for an entirely cometary composition \citep{castillo2023compositions}. Compared to the porous ice case with full radiogenic heating (Section~\ref{sec:resultporo}), the fraction of successful simulations rapidly decreases with less heating. Less available heat makes it difficult for the ice shell to warm up and convect or form an ocean except for the thickest hydrospheres. For the highest values of $\beta$, thick ice shells are able to spin down Eris, relying on the anelastic response to tides, but as noted above, such values may not be realistic; if $\beta\leq3\times10^{-11}$~Pa$^{-1}$ s$^{-0.25}$ is used as the criterion, all successful models include oceans. Thus, it is difficult to explain the current state of Eris with a low density rocky interior and reduced heating given a cometary composition. The thermal conductivity of rocks with organics may be decreased by an order of magnitude or increased by an order of magnitude for graphite \citep{castillo2023compositions}. The effects of organics on thermal conductivity may be important, but requires geochemical modeling that is out of the scope of this work. 

\section{Conclusions and implications}

Eris must have had a dissipative past based on its current orbital state, as \cite{nimmo2023internal} argued, and our thermal-orbital modeling finds that it is difficult to explain Eris's spun-down state without an ocean. Except for the cometary case with reduced heating, oceans make up $77-100$\% of successful simulations, which increases to $>98$\% across all cases when the Andrade $\beta$ parameter for ice is restricted to $\beta\leq3\times10^{-11}$~Pa$^{-1}$ s$^{-0.25}$. The formation of an ocean requires thick hydrospheres (for insulation) and moderate ice viscosities to prevent vigorous convection. Insulation by porous ice and gas clathrates or with antifreezes like ammonia can help an ocean form. Without insulation, oceans can last for 500 Myr to 2 Gyr but refreeze by the present day. Taking into account a porous insulating layer or by including antifreeze, oceans can readily survive to the present day.

Modes of insulation and their locations have a large impact. Porosity, gas clathrate, and other insulating layers can help maintain oceans for significant portions of Eris's history. Porosity is not as effective as surface clathrates, which are the best way to allow for both cases of oceans and convecting ice shells without oceans. Formation of gas hydrates near the surface requires a warm early history and differentiation, where a melted hydrosphere containing a large amount of volatiles released from water-rock reactions freezes, trapping gas in clathrates before they are lost to an atmosphere \citep{courville2023timing}. Further geochemical modeling of the early thermal history of Eris will help in understanding clathrate formation. Although not explored in this work, primordial oceans formed during differentiation and heating by short-lived radioisotopes might help to spin down Eris if maintained for a significant amount of time after Dysnomia's formation by insulation and/or tidal heating \citep{saxena2018relevance}. 

An expected consequence of a convecting ice shell (with or without an ocean) is relaxed surface topography due to its low viscosity. We would expect a lack of large scale features on Eris leading to a uniform distribution of surface volatiles, consistent with the observed shallow light curve \citep{bernstein2023synchronous, szakats2023tidally}, although near-infrared J and H-band observations suggest some surface heterogeneity in chemical composition and/or material properties \citep{szakats2023rotational}. A convecting ice shell should also cause the shape of Eris to conform to an equipotential \citep{nimmo2023internal}. 

A subsurface ocean on Eris, if formed, is likely freezing or frozen due to the gradual depletion of long-lived radioisotopes and lack of a significant additional heat source. A consequence of this is ocean pressurization, surface extension, and possibly cryovolcanism \citep{desch2009thermal}. Cryovolcanism can facilitate transport of material from the interior, which may leave observable signatures at the surface. D/H ratio measurements of surface methane ices by JWST suggest production by internal hydrothermal or metamorphic processes \citep{grundy2024measurement, glein2024moderate} which is potentially consistent with the presence of an ocean on Eris. Temperatures in the rocky core reach 150-400~K near the rock-hydrosphere interface where thermogenic methane can be produced, which then may be extracted from the core by hydrothermal circulation or diffuse outgassing and subsequently transported to the surface by cryovolcanism or ice convection \citep{glein2024moderate}. For the thermal evolutions shown in Figure~\ref{fig:EvolutionExample}, the top 60-80~km of the rocky core remain below the 400~K contour for most of Eris's history. This layer is thicker for smaller $\rho_{rock}$ and on average thinner if an ocean forms due to a steeper thermal gradient. Both cryovolcanism and convecting ice shells could refresh the surface of Eris, which may help explain the bright surface albedo \citep{sicardy2011pluto} and the apparently unevolved $^{13}$C/$^{12}$C ratio \citep{grundy2024measurement}. 

\printcredits

\section*{Acknowledgments}

The authors are grateful to the two anonymous reviewers whose constructive feedback helped improve this manuscript. This research was supported by NASA grant 80NSSC25K0199. 

\section*{Data availability}
Codes for the thermal-orbital evolution model are available upon request from R.A.

\appendix

\section{Rheology model}
\label{sec:appdxrheo}

This work uses several rheology models to relate the internal structure to the tidal response: Maxwell, Andrade $\beta$, and Andrade $\zeta$. These rheology models are described by the complex compliance $J^*$, which relates to the complex elastic modulus by $G^*=J^{*-1}$. Maxwell rheology is described by 
\begin{equation}
    J^* = \frac{1}{G} - i \frac{1}{\omega \eta},
\end{equation}
for elastic modulus $G$, forcing frequency $\omega$, and viscosity $\eta$. Andrade $\beta$ rheology is described by 
\begin{equation}
    J^* = \frac{1}{G} - i \frac{1}{\omega \eta} + \beta (i\omega)^{-n}\Gamma(n+1),
\end{equation}
where $\beta, n$ are Andrade parameters and $\Gamma$ is a gamma function. Andrade $\zeta$ rheology is described by 
\begin{equation}
    J^* = \frac{1}{G} - i \frac{1}{\omega \eta} + \frac{1}{G}\left(\zeta\frac{\eta}{G}i\omega\right)^{-n}\Gamma(n+1),
\end{equation}
where $\zeta, n$ are Andrade parameters. A comparison of dissipation $Q/k_2$ for different rheology models is shown in Supplementary Figure 6. 

\section{Supplementary data}
Supplementary material related to this article can be found online at 
\hyperlink{}{https://doi.org/10.1016/j.icarus.2025.116770}

\bibliographystyle{cas-model2-names}
\bibliography{bibliography}

\end{document}